\documentclass[conference]{IEEEtran}
\IEEEoverridecommandlockouts
\usepackage{cite}
\usepackage{amsmath,amssymb,amsfonts}
\usepackage{algorithmic}
\usepackage{verbatim}
\usepackage{graphicx}
\usepackage{textcomp}
\usepackage{xcolor}
\usepackage{booktabs}
\usepackage{multirow}
\usepackage{bm}
\usepackage[inline]{enumitem}
\usepackage{url}
\usepackage[breaklinks]{hyperref}

\usepackage{cleveref}
\def\BibTeX{{\rm B\kern-.05em{\sc i\kern-.025em b}\kern-.08em
    T\kern-.1667em\lower.7ex\hbox{E}\kern-.125emX}}
\begin{document}

\title{Significant Digits: Using Large-Scale Blockchain Data to Predict Fraudulent Addresses}

\author{\IEEEauthorblockN{Jared Gridley}
\IEEEauthorblockA{\textit{Rensselaer Polytechnic Institute} \\
Troy, NY, USA \\
gridlj@rpi.edu}
\and
\IEEEauthorblockN{Oshani Seneviratne}
\IEEEauthorblockA{\textit{Rensselaer Polytechnic Institute} \\
Troy, NY, USA \\
senevo@rpi.edu}}


\maketitle

\begin{abstract}
Blockchain systems and cryptocurrencies have exploded in popularity over the past decade, and with this growing user base, the number of cryptocurrency scams has also surged. Given the graphical structure of blockchain networks and the abundance of data generated on these networks, we use graph mining techniques to extract essential information on transactions and apply Benford's Law to extract distributional information on address transactions. We then apply a gradient-boosting tree model to predict fraudulent addresses. Our results show that our method can detect scams with reasonable accuracy and that the features generated based on Benford's Law are the most significant features.
\end{abstract}


\section{Introduction}
Over the past decade, the cryptocurrency ecosystem has exploded in every way, from market capitalization to user interaction. In 2013, there were just seven cryptocurrencies, with a market capitalization of about 1.5 billion USD. In March 2022, there were over 10,000 active cryptocurrencies with a total market cap of over 2 trillion USD \cite{source1}. With faster, cheaper, and more user-friendly blockchain technology,  cryptocurrencies have become more accessible to more people. The rising popularity of cryptocurrencies has piqued the interest of established financial institutions, with asset managers like BlackRock and J.P. Morgan Chase \& Co. disclosing virtual currencies on their balance sheets \cite{source2}. However, with such a rapidly growing environment, it becomes ripe for malicious users who seek to masquerade a rather useless smart contract as the next moonshot, and unfortunately, many people fall for these traps.

The rapid growth of cryptocurrency applications is paralleled by advancements in malicious tactics, particularly with Ponzi Schemes. For example, the most apparent scams on Bitcoin are Ponzi schemes where you send bitcoin to an address, and they promise to double it, often posing as a celebrity on social media \cite{trade_or_trick}. Ponzi schemes are often also characterized by a ``rug-pull event'' in which the orchestrator will disappear with a majority of the cash flowing through the scheme~\cite{bartoletti2018data}. However, rug-pull operations are not unique to Ponzi schemes, many other scams have similar events.

As Ethereum became popular, new scams appeared that took advantage of its smart contract technology. Bartoletti et al. analyzed the significant aspects that sparked the rise of Ponzi schemes with Ethereum's smart contracts. According to their analysis, the most critical factors for the rise in cryptocurrency scams are the anonymity among smart contract initiators, the immutable presence of malicious smart contracts, and the false sense of security many investors feel when interacting with smart contracts \cite{bartoletti2020dissecting}.

Unlike centralized fiat currencies backed by a government and law enforcement agencies, cryptocurrencies incur much more responsibility on the user. In 2021, it was reported that over \$14 billion was stolen in cryptocurrency scams, up 516\% from 2020, with 72\% of the stolen funds coming from Decentralized Finance (DeFi) protocols \cite{source5}. This sharp rise in scams makes it even more necessary for an identification system that tags scams before users engage in the next so-called ``moonshot.'' As more people use cryptocurrencies, more scammers will seize the opportunity to take advantage of new users in an unfamiliar ecosystem. If a user is caught in a Ponzi scheme, there is typically very little support from law enforcement agencies such as the FBI to help bring justice and retribution. With the permanency of transactions and the diversity of DeFi applications, a robust method for flagging potential scams is crucial for the financial security of blockchain-based applications.

\subsection{Challenges}
\label{subsec:challenges}

When building a classifier for cryptocurrency scams, there are two main challenges:
\begin{enumerate}[]
    \item \textbf{Data Sourcing:} We need a reliable source of scam addresses. To our knowledge, no source exists with such a comprehensive scam dataset. In many cases, smaller datasets exist for addresses associated with Ponzi schemes or phishing attacks but often rely on user reporting, meaning many scams are likely not included.
    
    \item \textbf{Scam Categorization:} The transaction patterns on a diverse chain such as Ethereum vary significantly. Transactions include a myriad of patterns with users, smart contracts, Maximal Extractable Value (MEV) bots~\cite{mev-bots}, and token contracts~\cite{token-contracts} operating on one chain. In many cases, some addresses have irregular patterns similar to scams but are innocent. We seek to avoid mislabeling an innocent address as a scam to encourage a more open, decentralized ecosystem.    
\end{enumerate}

\begin{figure}[!htbp]
    \centering
    \includegraphics[width=\columnwidth]{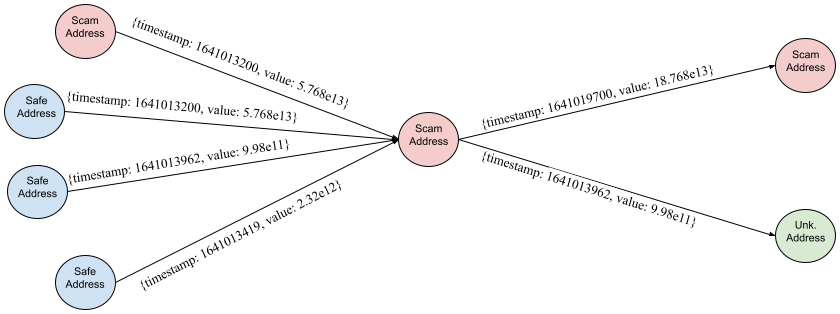}
    \caption{Transaction Subgraph for a Single Scam Address}
    \label{fig:transaction-subgraph}
\end{figure}

Scam addresses do, however, have distinctions that allow us to separate them from non-scam addresses. In particular, using obfuscation tools is common among scammers to try and clean their funds. It is important to note that while not addresses that use obfuscation applications like \emph{mixers} (or \emph{tumblers})~\cite{mixer} are scammers, many malicious users use these apps. An example sub-graph is depicted in \Cref{fig:transaction-subgraph}, showing that not all addresses connected to scam addresses is necessarily malicious.

While obfuscation techniques were most common on Bitcoin, where there are fewer exchanges to trickle funds through, they have quickly been adopted and built for Ethereum. In this work, we do not use a feature such as \emph{UsedObfuscationTool} because detecting ``coinjoins'' and ``mixers'' on a blockchain is a complicated area of research and development.

A particular trait we examine in this work comes from accounting fraud detection. When users make transactions, the first three significant digits follow specific, logarithm-based, non-uniform distributions. When malicious users hack into an account or convince users to send money, they often break these naturally occurring digit distributions for a uniform one \cite{uniform_scammers}. This distribution is characterized by ``Benford's Law for Anomalous Numbers''~\cite{benford_paper}. Analysis leveraging Benford's Law has even been admitted as evidence in criminal trials at all levels of court in the United States~\cite{radiolab_court}, making Benford's Law particularly interesting with blockchain scams because it is a method that is already accepted by regulatory bodies as credible evidence.


\section{Background}
\label{sec:background}

We outline some concepts pertinent to understanding our research contribution in this section.

\subsection{Phishing Schemes}
\label{subsec:phishing}

\emph{Phishing} is a social engineering attack that exploits system users to gain unauthorized access or steal funds. Traditional phishing attacks often consisted of a spam email or website that would deceive the recipient into giving up their passwords or personal information by impersonating a legitimate organization \cite{source6}. 

Phishing schemes on Ethereum have multiple avenues for attack. Attackers often target users directly by spreading phishing addresses and false Non-Fungible-Tokens (NFTs) or DeFi information on social media and chat rooms for other projects \cite{source7, source10}. Take the Bored Ape Yacht Club\footnote{\url{https://opensea.io/collection/boredapeyachtclub}}, for example. In 2021, Calvin Becerra, the owner of three Bored Ape NFTs, sent all three to another user address that claimed to be providing technical support. The scammer had stolen over \$1 million in NFT assets within this single transaction. While Becerra eventually got some of the money back, he had to transfer funds to the scammer before they were returned \cite{vice_bayc}.


A primary challenge with detecting phishing schemes on blockchain networks is that, in many cases, most malicious activity happens off the network. Social engineering tactics often target users with malicious emails and websites, making detecting phishing schemes especially challenging before a user's funds have been stolen. However, many researchers have investigated this problem. Wen et al. developed a phishing detection framework from on-chain transaction data and an adversarial attack framework to verify its robustness \cite{Wen_adversarial}. The idea of an adversarial method to improve the framework's robustness is significant, although the authors also emphasize the difficulty of developing phishing detection. 


\subsection{Ponzi Schemes}
\label{subsec:ponzi}

Ponzi schemes are often characterized by their advertisement as a High-Yield Investment Program (HYIP)\footnote{HYIPs usually advertise yields of more than 100\% per year to lure in victims and regularly use new investors' money to pay off older investors.}. They try to lure unsuspecting users with high-interest rates and the promise of high returns \cite{scam_classification, finacial_review_ponzi}. 
Much research has been done to detect Ponzi schemes that occur through malicious smart contracts, which we will refer to as ``Smart Ponzi Schemes.'' Many malicious users choose Smart Ponzi schemes because they can proliferate and bring in more money before being caught. 

There has been some effort towards educating investors about crypto scams from government agencies, like looking for registered investments with documented token information and strategies \cite{scam_classification, sec_ponzi}. However, given how quickly the crypto landscape changes, these sources often need more information, making an automated technique much more practical and effective. Such a solution can be implemented by leveraging machine learning techniques that classify new addresses as soon as they become active on the blockchain. 

\subsection{Benford's Law}
\label{sec:benfords_law} 

We utilize \emph{Benford's Law}~\cite{benford_paper} to create features for our machine learning classifiers. Benford's law is a natural phenomenon that maps the occurrence of first and second digits in many naturally occurring numerical sets to the base ten logarithms for each respective digit \cite{kessel_2020}. For example, the frequency of the occurrence of the number 1 would be calculated by:

\begin{equation}
    P(d) = log_{10}(1 + \frac{1}{d})
\end{equation}
\begin{equation}
    P(1) = log_{10}(1 + \frac{1}{1}) = 0.301...
\end{equation}

A Canadian-American astronomer, Simon Newcomb, first documented Benford's Law, who noticed the pattern by observing that in logarithm tables, the earlier pages (starting with 1 or 2) were much more worn than those that started with the latter digits \cite{newcomb_dicovery}. The law was later formalized by physicist Frank Benford who tested it on numerous naturally occurring datasets, including the surface area of 335 rivers, values of 140 physical constants, and weights of 1800 molecules \cite{benford_paper}.

While many naturally occurring datasets follow Benford's Law, many do not. For example, square roots and reciprocals of consecutive natural numbers, a list of local telephone numbers, and terminal digits in pathology data (due to rounding) violate Benford's Law~\cite{benford_violators}. 

General criteria for distributions that are expected to follow Benford's Law are given below~\cite{critera_Benfords}:
\begin{enumerate}
    \item Distributions where the mean is greater than the median and the skew is positive
    \item Numbers resulting from a combination (add/mult)
    \item Transaction-level data
\end{enumerate}

Benford's law has been used to detect fraud, particularly with fraudulent credit card transactions and applications in detecting money laundering and network intrusion. Each application of Benford's Law relies on the underlying distribution following Benford's Law and the fact that malicious actors tend to break this distribution and approach a more uniform one \cite{benford_scientific_data}. In sophisticated cases, it was found that many actors used transactions that followed Benford's Law for the first digits, but the illegal transactions still failed Benford's Law for the second digits. Previous works have shown that many aspects of cryptocurrency data follow Benford's Law  \cite{benford_scientific_data,critera_Benfords}.

\section{Problem Formulation}
\label{sec:problem}

In this work, we examine the transaction graph for Ethereum addresses and extract and transform the raw data into features used with various machine-learning classifiers. We focus on two primary research questions:
\begin{enumerate}
    \item To what extent does Benford's Law distinguish between fraudulent and legitimate users? 
    \item How can Benford's Law be used to build a more effective classifier for cryptocurrency Ponzi schemes? 
\end{enumerate}

Many previous methods extract smart contract code for the basis of their features. While the scams that operate on smart contracts grow much quicker, there are still scams that happen without a smart contract. We thus examine methods for predicting traditional Ponzi schemes and Smart Ponzi schemes. 

We also investigate the result of using features based on statistical fraud detection methods and, in particular, measuring the similarities between transactional value distributions and Benford's Law for first and second digits. 

By analyzing the Ethereum blockchain, we form a transaction graph G = (V, E) where the vertices V are addresses and edges E are the transactions between addresses. The edges hold transaction information, like the amount transferred, gas limit, and transaction timestamp. Graph mining techniques are then used to supplement the features derived from the distributions. 

We analyze online repositories of reported scam addresses to provide labels, $Y$, for the addresses in our graph where $Y = +1$ indicates a scam and $Y = -1$ indicates a non-scam. This graph is then used to extract features based on transaction statistics and distributions of the transaction values to then train a classifier.

\section{Data}
\label{sect:data}

Throughout this work, we investigated many sources to find comprehensive and reliable sources of Ethereum transaction data and reported scam data. Our dataset consisted of 1676 addresses with approximately 2.6 million transactions in total. This set of addresses consists of user activity, smart contracts, MEV Bots, and other DeFi applications.

\subsection{Blockchain Data Sourcing}
\label{subsec:tx_data}

We used an academic license to query the Amberdata API (\url{https://www.amberdata.io}) to collect information on cryptocurrency transactions. In addition to the raw on-chain data provided by Ethereum, they offer identifiers to transactions that belong to exchanges, DeFi applications, and transactions that span across different blockchains. We used Amberdata to get a much more comprehensive transaction history for our addresses and quickly sort out user addresses from smart contracts. 

\subsection{Class Label Sourcing}
\label{subsec:online}
 A particular challenge when creating the dataset was to ensure the integrity of the scam and non-scam data labels. 

For scam addresses, we sourced addresses from online repositories associated with other works in identifying crypto scams. Xia et al. developed a dataset of scam tokens that appeared on the Uniswap Exchange \cite{trade_or_trick}. In a later paper, Xia et al. developed a dataset of about 185 scam addresses across Bitcoin, Ethereum, and other blockchains and a similar dataset corresponding to scam web domains primarily used in phishing attacks \cite{char_crypto_exchange_scams}. In addition to these repositories, we used a GitHub repository created by Tomasz Nurkiewicz that aggregated news stories on significant crypto scams and the addresses associated with them \cite{nurkiewicz_2022}. Our most important source of scam addresses came from the Etherscan (\url{https://etherscan.io}) tagging system. Their tagging system (\url{https://etherscan.io/labelcloud}) identifies 564 different labels ranging from the addresses and smart contracts associated with Uniswap to addresses associated with reported phishing attacks. Etherscan has a free API that provides access to these labels \cite{etherscan}.

When gathering the non-scam addresses, we similarly used Etherscan labels to pull addresses for trusted smart contracts. In particular, we pulled addresses associated with Uniswap, Aave, Compound, and OpenSea. While these applications are considered reliable, many addresses have thousands of transactions. So to account for user addresses, we looked at addresses verified by DeFi applications, particularly on OpenSea\footnote{\url{https://opensea.io}} and Axie Infinity\footnote{\url{https://axieinfinity.com}}. The remaining non-scam addresses came from pulling addresses that had traded on a set of blocks in March 2022 and checking them against user-reported scams on ScamAlert\footnote{\url{https://scam-alert.io}}. 

\subsection{Feature Extraction}

To get the features we used to train our classifiers, we extracted the transaction graph for each address and then used that to generate a statistical representation of the transaction graph. We examined the number of transactions, unique addresses, values for gas limits, and value transferred. Each feature was broken down between incoming and outgoing transactions, and the gas limit and value metrics were represented by their mean, median, and standard deviation values.
We used this breakdown of the transaction graph to generate our features because, in previous work that looked to classify scam tokens~\cite{trade_or_trick}, there was a similar representation of the transaction graph worked well in their token classifiers. We modified it by adding the gas limits and median to the feature set. 
These features were then supplemented with the Chi-Squared and KS test values for the first and second digits to quantify their fit with Benford's Law.

\section{Methodology}
\label{sec:methodology}

For this work, we broke down our investigation into two parts. The first is testing whether Benford's Law fits cryptocurrency data for legitimate and scam-labeled data. The second part is building a series of classifiers for scam addresses based solely on the transaction graph. 

\subsection{Measuring Fit with Benford's Law}
\label{subsec:measuring_fit}

To measure the fit with Benford's Law, we first separated the addresses by their scam and non-scam labels and using two metrics, the Chi-Squared~\cite{greenwood1996guide} and Kolmogorov–Smirnov (KS)~\cite{massey1951kolmogorov} tests, to quantify the similarities. The Chi-Squared test is recommended for distributions with many sample data. However, since not all of the addresses in our dataset have many transactions, we also consider the KS test because it has been shown to better account for the minor differences in the distributions \cite{kolmogorov-smirnov}. We used both features in our classifier but later found that the KS test is not significant in any of our classifiers.

\subsection{Building Classifiers}
\label{subsec:classifiers}

For this investigation, we considered five machine-learning classification methods. We used: (i) Logistic Regression~\cite{kleinbaum2002logistic}, (ii) Random Forest~\cite{breiman2001random}, (iii) Support Vector Machine (SVM)~\cite{noble2006support}, (iv) Decision Tree~\cite{safavian1991survey}, and (v) LightGBM~\cite{ke2017lightgbm}. 
LightGBM is a gradient-boosting framework that uses tree-based learning algorithms. Microsoft initially developed it, and is now an open-source tool~\cite{lightgbm-github}.

We randomly split our data, with 20\% of it being split as test data and 80\% as training. The training data is further split, with 15\% being validation data and the rest as training data.

\begin{table}[!htpb]
\centering
\label{tbl:data-composition}
\caption{Data Splits\vskip 3pt }
\begin{tabular}{|c|ccc|}
\hline \bf & \bf Training & \bf Validation  & \bf Testing\\ \hline
Non-Scam & 59.4\% & 14.8\%  & 18.4\% \\
Scam& 4.6\%  & 1.2\%  & 1.6\% \\
\hline
Total & 64.0\% & 16.0\% & 20\%\\
\hline
\end{tabular}
\end{table}


\section{Results}
\label{sec:results}

\subsection{Cryptocurrency and Benford's Law}
\label{subsec:crypto_and_benfords_law}

\begin{figure}[thb]
    \centering
	\includegraphics[width = \columnwidth]{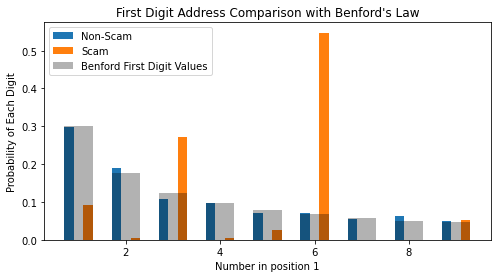}
	\caption{Examining a scam and non-scam address}
	\label{fig:benford-single-addrs}       
\end{figure}

When investigating the distribution of transactions in relation to Benford's Law, we found that the scam addresses had a clear divergence in many cases. In Figure \ref{fig:benford-single-addrs}, we compare two addresses, each with a similar number of transactions (the scam address had 1404, and the non-scam address had 1426). The non-scam address (blue) follows Benford's law quite closely, whereas the scam address (orange) does not fit Benford's Law at all, which is naturally not the case for all scam addresses, with some addresses having much more subtle differences. 

While many scam addresses had very little correlation with Benford's Law, we found that when examining the scam transactions, the distribution mapped much closer to Benford's law. However, there are still discrepancies with the digits 1 and 5 primarily, which can be seen more clearly in \Cref{fig:benford-first-digits} below, where we compare all the transactions in each category to Benford's law. 
\begin{figure}[thb]
    \centering
	\includegraphics[width = \columnwidth]{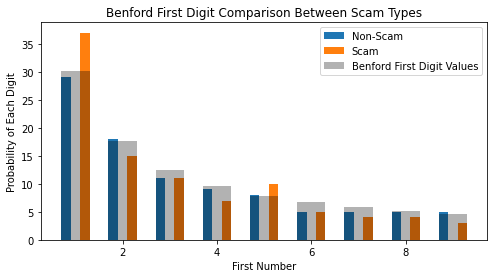}
	\caption{Benford's Law on First Digits on all transactions by class}
	\label{fig:benford-first-digits}       
\end{figure}

In mapping the distribution of the digits in the second position, we found that both categories had more occurrences of the digit 0 than Benford's Law for second digits, but the scam category was still significantly higher than the non-scam. This mapping caused the non-scam category to follow Benford's Law much more closely than the scam category as it had a smaller margin in the occurrences of 0, so other digits were not as divergent from Benford's Law. 

Using the Chi-Squared and KS tests on all scam/non-scam transactions, we then measured the digit distributions fit with Benford's Law. We found that both distributions fit Benford's Law for the first digit quite well, although the non-scam transactions still had a closer fit. For the second digits, neither distribution fit as well as the first digits, but the non-scam transactions had a significantly closer fit than the scam transactions, as seen in \Cref{tab:_benford_tests} with the Chi-Squared test in particular. For individual addresses, we found many more scam addresses with a higher Chi-Squared test value. The mean for the first digit Chi-Squared values among all the scam addresses was 1.37, compared to 1.01 for non-scam addresses. This gap significantly widens when looking at the second-digit Chi-Squared test values. The scam addresses had an average of 3.29, whereas the non-scam addresses averaged 1.13. This further indicates a distinguishing feature between the two classes.


\begin{table}[thb]
\centering
\caption{Benford's Law Fit Results\vskip 3pt }
\label{tab:_benford_tests}
\begin{tabular}{cc|cc}
\hline &  & \bf Chi-Squared  & \bf KS Test \\ \hline\hline
1st Digit   & Scam      & 0.0388  & 0.333\\
            & Non-Scam  & 0.0047  & 0.222 \\
            \hline
2nd Digit   & Scam      & 0.5854  & 0.700\\
            & Non-Scam  & 0.0997  & 0.400\\
\hline
\end{tabular}
\end{table}

The Chi-Squared and KS tests clearly distinguish between the distributions for scam and non-scam transaction values. Both metrics supplement the statistical transaction features in training the classifiers. However, we can already predict that the second digit distributions will be a more effective separating feature than the 1st digit. Further, the Chi-Squared test will be a better separator than the KS test as it is more sensitive to the differences between two distributions.

The results in Table \Cref{tab:_benford_tests} can help to answer our first research question on the effectiveness of Benford's Law at separating between a scam and non-scam cryptocurrency addresses. The results from the first digit distributions show a noticeable separation between scam and non-scam; however, it is a considerably slim margin. The second digit distributions show a much more significant margin between scam and non-scam, which draws us to the conclusion that Benford's Law for Second Digits provides a helpful distinguishing feature, whereas the first digit distribution is not very effective. This result is further reinforced by our results in the next section, which shows that the second-digit features rank much higher in importance than the first-digit features. 

\subsection{Classifiers with Benford's Law Features}
\label{subsec:classifiers_and_benfords_law}

As seen in Table \ref{tab:classifier_results}, the LightGBM model performed better than the other methods examined, which was expected and followed our results with the validation dataset. The decision tree with Adaboost~\cite{safavian1991survey} was the second closest in correctly classifying the scam addresses (recall), but it was limited by its misclassification of the non-scam addresses (precision). The LightGBM model~\cite{ke2017lightgbm} significantly outperformed the decision tree on the test data. 

\begin{figure}[thb]
    \centering
	\includegraphics[width = \linewidth]{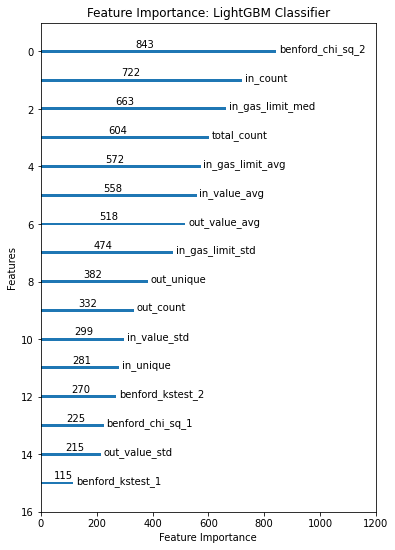}
	\caption{Feature Importances for the LightGBM Model}
	\label{fig:lightGBM-importances}       
\end{figure}

With the Support Vector Machine and the Logistic Regression Model, the classifier tended to fall into the trap of classifying everything as non-scam. We expected these models to perform poorly, and many features were similar to non-scam addresses, and their poor performance also likely resulted from the dataset's class imbalance. When looking at feature importance for the non-tree-based model, we found that the model only used 3-4 primary features for classification, always with a feature based on Benford's Law for second digits. The LightGBM model, however, appears to have a less skewed feature ranking, as seen in Figure \ref{fig:lightGBM-importances}, which is expected, given that LightGBM is designed to build more robust models.

When examining the feature importance for each model, it was found that the Chi-Squared measurement for the second digit was considered an essential feature in the logistic regression, random forest, and LightGBM models and was the second-most important feature in the decision tree model. In the SVM, it did not rank high in terms of importance. However, as expected, the SVM model was the worst-performing among the methods tested. In the LightGBM model, it was an essential feature, which is seen more clearly in Figure \ref{fig:lightGBM-importances}. The LightGBM indicates that Benford's Law is an effective way to separate the scam from the non-scam. We tested the effectiveness of classifiers without Benford's Law features. Those results are discussed in the following section.

\begin{table*}[thb]
\centering
\caption{Benford's Law Feature Classifier Results\vskip 3pt }
\label{tab:classifier_results}
\begin{tabular*}{\textwidth}{cc|ccccc}
\hline & & \bf Logistic Regression  & \bf Random Forest & \bf Support Vector Machine & \bf Decision Tree w/Adaboost & \bf LightGBM\\ \hline 
\hline
&Macro Avg Precision  & 0.5851     & 0.8990 & 0.4629 & 0.7891 & 0.9544 \\
Without &Macro Avg Recall      & 0.5127    & 0.6693 & 0.4981 & 0.8249 & 0.7916 \\
Benford &Macro Avg F1-Score  & 0.5073     & 0.7282 & 0.4799 & 0.8056 & 0.8519 \\
Features&Macro Avg Accuracy  &  0.5126  & 0.6693    & 0.4984 &  0.7732 & \textbf{0.7916} \\
& Accuracy    & 0.9127    & 0.9408 & 0.9155 & 0.9296 & 0.9634 \\
\hline
\hline

&Macro Avg Precision  & 0.8568     & 0.9794 & 0.5851 & 0.8164 & 0.9852 \\
With &Macro Avg Recall      & 0.6678    & 0.7586 & 0.5126 & 0.7743 & 0.8095 \\
Benford &Macro Avg F1-Score  & 0.7216     & 0.8304 & 0.5074 & 0.7935 & 0.8749 \\
Features &Macro Avg Accuracy    & 0.6678    & 0.7586 & 0.5126 & 0.8153 & \textbf{0.8966} \\
& Accuracy    & 0.9380    & 0.9606  & 0.9172 & 0.9493 & \textbf{0.9831}\\
\hline
\hline
\end{tabular*}
\end{table*}

Interestingly, in Figure \ref{fig:lightGBM-importances}, the KS test ranked relatively low for both the first and second digits, likely due to the nature of scam data. Most scams in the dataset had many transactions, resulting from the fact that many were operated through smart contracts and could thus grow quicker. This phenomenon is seen clearly in Table \ref{tab:_benford_tests} as there is a considerable gap between the scam and non-scam results, but both performed poorly. However, according to the feature ranking results, the Chi-Squared results are essential to distinguish between scam and non-scam addresses for second digits.

\subsection{Classifiers without Benford's Law Features}
\label{subsec:classifiers_no_benfords}

We also trained the classifiers without the features related to Benford's Law to measure the improvement or deterioration of the Benford's Law features. We found that nearly every model performed worse with lower precision, recall, and F1-score than with Benford's Law features. The exception was the decision tree with Adaboost, which had an overall lower accuracy without the Benford's Law features, resulting from a lower precision but a higher recall. From Table \ref{tab:classifier_results}, we can see that the accuracy with Benford's Law features increased by about two percentage points on average, with the macro average accuracy increasing by 0.105 in the LightGBM model and 0.0421 in the decision tree model, which suggests that features related to Benford's Law can help with over-fitting as improving the macro average results from accuracy improvement in each class. 

These results help to answer our second research question on the effectiveness of Benford's Law at classifying addresses. With the improvement in both macro average accuracy and weighted average accuracy from the addition of Benford's Law features, we can conclude that Benford's Law is very effective as a training feature for classification.

\section{Related Work}
\label{sec:related-work}

Many academic and commercial solutions have been developed to identify phishing attacks. Abdelhamid et al. proposed a multi-label classification method to tackle phishing websites by extracting correlations in website features and similarity in URLs in particular \cite{source8}. Zouina et al., on the other hand, extracted features from website URLs and trained an SVM to classify phishing scams, achieving an accuracy score of 0.956 \cite{source9}. Many of these detection systems rely on features not apparent from the transaction graph, so the assessment of an address alone is limited. For this reason, most of our scam data in this paper come from Ponzi schemes, as they are scams where most activity happens on the blockchain. 

In detecting malicious smart contracts, Chen et al. proposed a method that examines the bytecode of the smart contract to extract features for classification through a dual-ensemble method to address the class imbalance problem \cite{chen_ponzi_scheme_eth}. It was shown to perform well and detect smart Ponzi schemes before they attract a significant victim base \cite{chen_ponzi_scheme_eth}. While this approach is great for tackling the most significant and damaging Ponzi schemes on Ethereum, those that operate without a smart contract can slip through the cracks. This work examines transactional data (not bytecode) of addresses operating on Ethereum, including smart contracts, MEV bots, and human users. 
As many of the models shown in this paper likely struggled with class imbalance, using a dual-ensemble model proposed by Chen et al. \cite{chen_ponzi_scheme_eth} would be an exciting avenue for further research.

Specifically, with Bitcoin, much of the research takes a graphical approach to feature extraction when examining address-based Ponzi schemes. Address-based schemes resemble traditional Ponzi schemes of sending money to another person's address. Bartoletti et al. proposed a set of features that focused on the lifetime and activity of Bitcoin addresses before applying three different classifiers: Repeated Incremental Pruning to Produce Error Reduction (RIPPER)~\cite{cohen1995repeated}, Bayes Network~\cite{friedman1997bayesian}, and a Random Forest~\cite{breiman2001random}, with varying cost constraints. The random forest approach yielded the best results across all cost configurations. When crafting their dataset, they considered the skewed distribution of Ponzi scheme addresses to legitimate addresses, testing on a dataset of 32 Ponzi schemes and 6000 legitimate addresses \cite{bartoletti2018data}. We apply similar graph-based feature extractions with the exception of the address lifetime. The features used in this work focus on measuring the frequency and value of transactions and gas limits, then supplementing with features measuring fit with Benford's Law. 


Within the Ethereum ecosystem, Xia et al. proposed a method of detecting scam tokens on the Uniswap decentralized exchange \cite{trade_or_trick}. They generated their dataset by looking at tokens with identical tickers to legitimate tokens and reported scam tokens from Etherscan, then applying a Guilt-By-Association expansion on the creators of these scam tokens to see which other tokens they created, further classifying them as scams. They queried their data from The Graph (\url{https://thegraph.com/hosted-service/}) and extracted features on both the tokens themselves and early investors before training many different machines learning classifiers to determine the best performing model. The random forest model performed best with precision, recall, and an F1 score all-around 0.96. They recognize the particular challenge of ground-truth labeling. As their model predicts scams, they must investigate the newly unclassified addresses, often finding suspicious activity but not enough to confidently say it was a scam. While this paper focused on the Uniswap token specifically, we found that the features they used to train their model were very comprehensive and used similar features when designing our model. By contrast, our work focuses on classifying all address entities on Ethereum rather than a specific exchange.

Much previous work has focused on Ponzi schemes that operate with smart contracts, classified as ``Smart Ponzi Schemes.'' Many malicious users choose Smart Ponzi schemes because they can proliferate and bring in more money before being caught. Chen et al. proposed a method that looks at the bytecode of the smart contract to extract features before training an XGBoost classification model \cite{chen_ponzi_scheme_eth}. Chen et al. furthered their work on Smart Ponzi Schemes with a novel dual-ensemble classification method focused on overcoming the class imbalance problem. It was shown to perform well and detect Ponzi schemes before they attract a significant victim base \cite{chen_eth_ponzi_exploit}. These approaches are great for tackling the most extensive and damaging Ponzi schemes, which comprise most of the Ponzi schemes on Ethereum. However, many smaller Ponzi schemes without a smart contract can slip through the cracks.

An exciting field within blockchain security is Graph Neural Networks. Shen et al. developed a neural network framework to infer the identity of users on a network by examining a subgraph of the user's activity \cite{shen2021identity}. Their method significantly improved baseline models, which they attribute to a deeper convolution layer and more compelling features. Further, Liu et al. developed a hyperbolic graph neural network to identify the hierarchical structure of subsection of the Ethereum ecosystem \cite{liu2019hyperbolic}. Their method was able to identify the most influential entities on the network in accordance with the address data compiled by Etherscan. Although many works focus on using graph neural networks for identity classification, their applications to fraud detection are an exciting avenue for further research.

\section{Conclusion and Future Work}
\label{sec:future-work}

From our research into related works (\Cref{sec:related-work}), this is the first paper to examine the use of Benford's Law to predict scams in cryptocurrencies. 
With recent actions by the US Department of Justice to bring charges against fraudulent cryptocurrency actors, Benford's Law can prove to be a crucial piece of evidence in investigations as it has previously been admitted as evidence in local, state, and federal courts \cite{matthews_2022, radiolab_court}. Thus, methods that use Benford's Law to classify scams have been used as evidence for legal action in the United States.

Further, financial scams will become a more critical research problem as cryptocurrencies become more widely used. We demonstrated the importance of a classical fraud detection method in the new financial ecosystem powered by blockchain. We created a gradient-boosted tree model using the labeled scam data and the LightGBM library. The experimental results indicate that Benford's Law distinguishes between scam addresses and non-scam addresses, and those metrics involving Benford's Law for second digits are a vital feature for classification. The most significant result of our method is that it relies solely on blockchain transaction data. By examining on-chain and internal transactions, our model can detect scams that operate with or without smart contracts or bots, spanning the range of attack sophistication.

Separating the classification task into two may prove beneficial for more accurate detection of the distinctions in behavior between smart Ponzi schemes and traditional Ponzi schemes. Using a data source, such as Amberdata, that can separate smart contracts from addresses would be helpful in this direction, as you could use a more robust code analysis method to reinforce a model targeting traditional schemes. 

Another area for further research lies in getting a better metric to match the fit with Benford's Law. While the Chi-Squared method performs exceptionally well with larger sample sizes, it is limited by sample size. So with very few addresses, a better metric could yield a better feature set, resulting in a better-performing classifier. Conversely, the Kolmogorov-Smirnov test proved to be ineffective in classification. A robust metric for comparing small and large samples to Benford's Law would be central to improving its applicability to detecting fraudulent transactions concerning cryptocurrencies.

\section*{Acknowledgements}

We acknowledge the support from Amberdata.io for giving us a free academic license to access their API.
We also acknowledge the support from National Science Foundation Industry–University Cooperative Research Centers (NSF IUCRC) Center for Research toward Advancing Financial Technologies (CRAFT) research grant for this research.

\bibliographystyle{ieeetr}
\bibliography{citations}
\end{document}